\documentclass[twocolumn,prc,aps,floats,floatfix,showpacs]{revtex4}
\input{epsf.tex}
\newcommand{\lapprox}{\raisebox{-0.5ex}{$\ 
\stackrel{\textstyle<}{\textstyle\sim}\ $}}

\begin{document}
\preprint{SWAT/07/501}
\title{CRITICAL FLAVOR NUMBER IN THE THREE DIMENSIONAL THIRRING MODEL}

\author{Stavros Christofi$^a$, Simon Hands$^{b}$ and Costas Strouthos$^c$}

\affiliation{$^a$Frederick Institute of Technology, CY-1303 Nicosia, Cyprus}

\affiliation{$^b$Department of Physics, Swansea University,
Singleton Park, Swansea SA2 8PP, U.K.}

\affiliation{$^c$
Institute of Chemical Sciences and Engineering, \'Ecole Polytechnique
F\'ed\'erale de Lausanne, 1015 Lausanne, Switzerland.}


\begin{abstract}  

We present results of a Monte Carlo simulation of the three dimensional
Thirring model with the number of fermion flavors $N_f$ varied between 2 and
18. By identifying the lattice coupling at which the chiral condensate
peaks, simulations are be performed at couplings $g^2(N_f)$ corresponding
to the strong coupling limit of the continuum theory. The chiral symmetry
restoring phase transition is studied as $N_f$ is increased, 
and the critical number of flavors
estimated as $N_{fc}=6.6(1)$. The critical exponents measured at the transition
do not agree with self-consistent solutions of the
Schwinger-Dyson equations; in particular there is no evidence for the
transition being of infinite order. Implications for the critical flavor number
in QED$_3$ are briefly discussed.
\end{abstract}


\maketitle

The study of quantum field theories in which the ground state shows a 
sensitivity to the number of fermion 
flavors $N_f$ is intrinsically interesting. According to certain 
approximate solutions of Schwinger-Dyson equations (SDEs), in $d=3$ spacetime
dimensions both quantum electrodynamics (QED$_3$) and 
the Thirring model display this phenomenon.
Both models have been proposed as effective theories describing different
regions of the cuprate phase diagram, Thirring describing the superconducting
phase, while QED$_3$ supposedly describes the
non-superconducting ``pseudogap'' behaviour seen in the underdoped 
regime \cite{herbut,cuprate}.
The Thirring model is a theory of fermions interacting via a current 
contact interaction:
\begin{equation}
{\cal L}=\bar\psi_i(\partial{\!\!\! /}\,+m)\psi_i
+{g^2\over{2N_f}}(\bar\psi_i\gamma_\mu\psi_i)^2,
\label{eq:L}
\end{equation}
where $\psi_i,\bar\psi_i$ are four-component spinors, $m$ is a 
parity-conserving bare mass, and the index $i$ runs over $N_f$ distinct
fermion flavors. 
In the chiral limit $m\to0$ the Lagrangian (\ref{eq:L}) shares the same  global
U(1) chiral symmetry $\psi\mapsto e^{i\alpha\gamma_5}\psi,\,
\bar\psi\mapsto\bar\psi e^{i\alpha\gamma_5}$ as QED$_3$.
Since the coupling $g^2$ has mass dimension $-1$, naive
power-counting suggests that the model is non-renormalisable.
However~\cite{parisi,gomes,hands1}, an
expansion in powers of $1/N_f$, rather than $g^2$, is exactly renormalisable
and the model has a well-defined continuum limit corresponding to a UV-stable
fixed point of the renormalization group (RG).
The $1/N_f$ expansion may not, however, describe 
the true behaviour of the model
at small $N_f$. Chiral symmetry breaking, signalled by a condensate
$\langle\bar\psi\psi\rangle\not=0$, is forbidden at all orders
in $1/N_f$, and yet is predicted by a self-consistent SDE approach 
\cite{gomes,hong,itoh,sugiura}. 
SDEs solved in the limit $g^2\rightarrow \infty$ \cite{itoh} 
show that chiral symmetry is spontaneously
broken for $N_f < N_{fc} \simeq 4.32$,
close to certain predictions of $N_{fc}$ 
for non-trivial IR behaviour in QED$_3$ \cite{kondo}.
Based on these results, at $N_f=N_{fc}$ the model is expected to 
undergo an infinite order or conformal
phase transition, originally discussed by Miranskii {\it et al} 
in the context of quenched QED$_4$ \cite{miranskii,conformal}.
Using different sequences of truncation, however, other SDE approaches
have found values of $N_{fc}$ ranging between 3.24 \cite{gomes}
and $\infty$ \cite{hong}. 

Since for $N_f<N_{fc}$
the chiral symmetry breaking transition corresponds to a UV-stable
RG fixed point at a critical $g_c^2$, 
which can be analysed using finite volume scaling techniques,
study using lattice Monte Carlo
simulation has proved possible with relatively modest computer resources 
\cite{hands2,hands3,hands4,hands5}. It has been shown that
for $N_f \leq 5$
the model is in the chirally broken phase at strong enough 
coupling $g^2<\infty$, providing a lower bound on $N_{fc}$,
and the critical exponents found to vary with $N_f$.
This should be contrasted with numerical studies of
QED$_3$: in this case the phase transition in the continuum limit
at $N_f=N_{fc}$ is an IR-stable
fixed point (see, eg. \cite{appelquist,maris}), and SDE predictions for the
separation of scales parametrised by the dimensionless
quantity $\langle\bar\psi\psi\rangle/\alpha^2$ are so small that 
enormous lattices are required to establish whether the model
with a particular $N_f$ lies
in the broken or symmetric phase \cite{qed3b}. A determination of $N_{fc}$ for
QED$_3$ (whose value may have profound implications for the cuprate phase
diagram \cite{cuprate}) by purely numerical means currently appears very
difficult. 

As already noted, Thirring and QED$_3$ share the same global symmetries;
moreover in the strong coupling limit the $1/N_f$ expansion predicts the 
existence of a massless spin-1 $\psi\bar\psi$ bound state in the Thirring
spectrum \cite{hands1,itoh}. 
The RG fixed point of both models is therefore a theory of massless
fermions interacting via massless vector exchange between conserved currents.
It is therefore plausible that the two models have the same $N_{fc}$,
and that simulation of the Thirring model is the most effective means to
determine its value.
Here, we present results based on numerical 
simulations with $N_f=2,\ldots,18$ 
in an effort to determine $N_{fc}$ with unprecedented precision, exploiting a
strategy of identifying the value of the lattice coupling $g^2$ corresponding to
the strong coupling limit of the continuum theory.

The lattice action we have used is based on \cite{hands2} and is as follows:
\begin{eqnarray}
      S &=& \frac{1}{2} \sum_{x\mu i} \bar\chi_i(x) \eta_\mu(x)
        (1+iA_\mu(x)) \chi_i(x+\hat\mu)
        - \mbox{h.c.} \nonumber \\
          & & + m \sum_{xi} \bar\chi_i(x) \chi_i(x) +
        \frac{N}{4g^2} \sum_{x\mu} A_\mu(x)^2 
\label{eq:lat}
\end{eqnarray}
where $\chi,\bar\chi$ are staggered fermion fields and the flavor index $i$ runs
over $N$ species. We have introduced an auxiliary real-valued link field
$A_\mu$, superficially resembling a photon field,
so that the lattice action can be expressed as a bilinear form. The formulation
is not unique; Gaussian integration over the $A_\mu$ in (\ref{eq:lat})
results in a lattice action resembling the continuum form (\ref{eq:L}) in that 
all interactions remain of the form $\bar\chi\chi\bar\chi\chi$ for arbitrary
$N$, but some of which when re-expressed in terms of continuum-like Dirac
spinors are of non-covariant form. As argued in \cite{hands3}, in the $1/N_f$
expansion these unwanted contributions are probably
irrelevant, but more care may be
needed when discussing the UV fixed-point theory.

For $d=3$ $N$ staggered fermion species describe $N_f=2N$
continuum flavors.
We employed the Hybrid Monte Carlo (HMC) algorithm to simulate even $N_f$
\cite{hands3} and  
the Hybrid Molecular Dynamics (HMD) algorithm for odd or non-integer 
$N_f$ \cite{hands5}. 
For the HMD simulations we used a small enough 
fictitious time step $\Delta \tau \leq 0.0025$
to ensure that the $O(N^2 \Delta \tau^2)$ systematic errors 
in the molecular dynamics steps 
were smaller than the statistical errors of the various observables. 

\begin{figure}[htb]
\begin{center}
\vspace*{-.2cm}
\epsfxsize=3.2in
\epsfbox{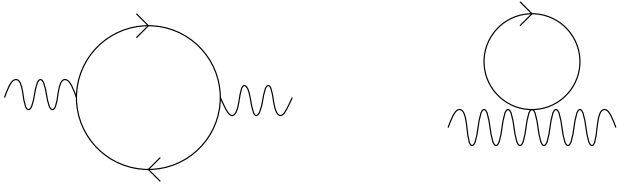}
\end{center}
\vspace{-.3cm}
\caption{
\label{fig:vacpol}
Leading order vacuum polarisation in lattice QED.}
\vspace{-.35cm}
\end{figure}
Next we review the discussion of \cite{hands3} 
regarding the non-conservation of the 
vector current in the lattice Thirring model. 
The leading order $1/N$ quantum corrections to the photon propagator in
lattice QED$_3$ are sketched in Fig.~\ref{fig:vacpol}.
The second diagram, arising from the gauge-invariant form
$\bar\chi_x e^{iA_{\mu x}}\chi_{x+\hat\mu}$,
is peculiar to the lattice regularisation, but is required to ensure
transversity of the vacuum polarisation tensor, ie:
\begin{equation}
\sum_\mu\left[\Pi_{\mu\nu}(x)-\Pi_{\mu\nu}(x-\hat\mu)\right]=0. 
\label{eq:transverse}
\end{equation}
For the $A_\mu$ propagator in
the lattice Thirring model (\ref{eq:lat}) however, the second diagram 
is absent, and the transversity condition (\ref{eq:transverse}) violated by a
term of $O(a^{-1})$. Transversity of $\Pi_{\mu\nu}$ is crucial to the
renormalisability of the $1/N_f$ expansion; fortunately, 
the impact of the extra divergence can be absorbed  by a wavefunction
renormalisation of $A_\mu$ and a coupling constant renormalisation
\begin{equation}
g_R^2={g^2\over{1-g^2J(m)}},
\end{equation}
where $J(m)$ is the value
of the integral contributed by the second diagram in Fig.~\ref{fig:vacpol}.
The physics described by continuum $1/N_f$ perturbation theory occurs for
the range of couplings $g_R^2\in[0,\infty)$, ie. for $g^2\in[0,g_{\rm lim}^2)$
where to leading order in $1/N$ $g_{\rm lim}^2=\frac{3}{2}$ for $m=0$.
The strong coupling limit is therefore recovered at $g^2=g^2_{\rm lim}$.
For $g^2>g^2_{\rm lim}$ the auxiliary propagator becomes negative, and the
lattice model no longer describes a unitary theory.

As already discussed, chiral symmetry breaking 
is absent in large-$N_f$ calculations.
It may well be that the value of the second diagram in Fig.~\ref{fig:vacpol} 
is considerably altered  in the chirally broken vacuum expected for
$N_f<N_{fc}$. In this study we use lattice simulations to estimate the 
value of $g^2_{\rm lim}$ as a function of $N_f$.
\begin{figure}[htb]
\begin{center}
\vspace*{-.8cm}
\epsfxsize=3.5in
\epsfbox{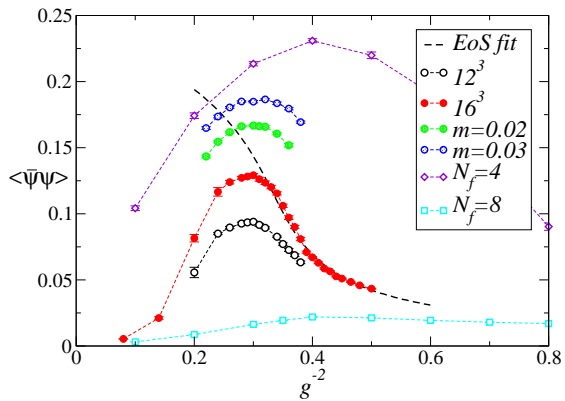}
\end{center}
\vspace{-.6cm}
\caption{
\label{fig:psibpsivsb}
Chiral condensate $\langle\bar\psi\psi\rangle$ vs. $g^{-2}$. $N_f=6$, $m=0.01$
on a $16^3$ lattice unless otherwise stated. The dashed line denotes a fit of
the $16^3$, $N_f=6$, $m=0.01$ data 
to an EoS of the form (\ref{eq:eos}) taking finite
volume scaling into account.}
\vspace{-.25cm}
\end{figure}
Fig.~\ref{fig:psibpsivsb} shows $\langle\bar\psi\psi\rangle$ 
data as a function of
$g^{-2}$ for various $N_f$, $m$ and lattice volumes. The condensate is
non-monotonic, showing a clear maximum for $N_f=6$ flavors at $g^{-2}\simeq0.3$.
The peak position at $g^2=g^2_{\rm max}$, 
unlike the value of the condensate itself,
is independent of both lattice volume and fermion mass, indicating that its
origin is at the UV scale. Since in an orthodox description of chiral symmetry
breaking $\vert\langle\bar\psi\psi\rangle\vert$ is expected to increase with 
the strength of the interaction, we interpret the peak as the point where 
unitarity violation sets in, ie. $g^2_{\rm max}\approx g^2_{\rm lim}$, and hence
identify the peak with the location of the strong coupling limit.

The figure also shows that $g^2_{\rm max}$ depends on $N_f$ (Cf. Fig.~3 of
Ref.~\cite{hands3}). For $N_f<N_{fc}$, it is possible to fit condensate data
taken at $g^2<g^2_{\rm max}$ to an equation of state (EoS) of the 
form~\cite{hands3}
\begin{equation}
m=A(g^{-2}-g_c^{-2})\langle\bar\psi\psi\rangle^p
+B\langle\bar\psi\psi\rangle^\delta,
\label{eq:eos}
\end{equation}
which describes a continuous transition in the limit $m\to0$ at
$g^2=g_c^2$ to a chirally symmetric phase, characterised by critical exponents
$\delta$ and a ``magnetic'' exponent $\beta=(\delta-p)^{-1}$. On a dataset taken
on $12^3,\ldots,32^3$ taking finite volume scaling into account
we have found the best fit for $N_f=6$ given by
$g_c^{-2}=0.316(1)$, $\delta=5.75(13)$, $p=1.18(2)$, to be compared with
corresponding quantities tabulated for smaller $N_f$ in \cite{hands5}. Since
$g_c^2\lapprox g^2_{\rm max}$, the EoS fit for $N_f=6$, 
shown for the $16^3$ data by a dashed line in
Fig.~\ref{fig:psibpsivsb}, is of borderline 
credibility.

\begin{figure}[htb]
\begin{center}
\vspace*{-.8cm}
\epsfxsize=3.5in
\epsfbox{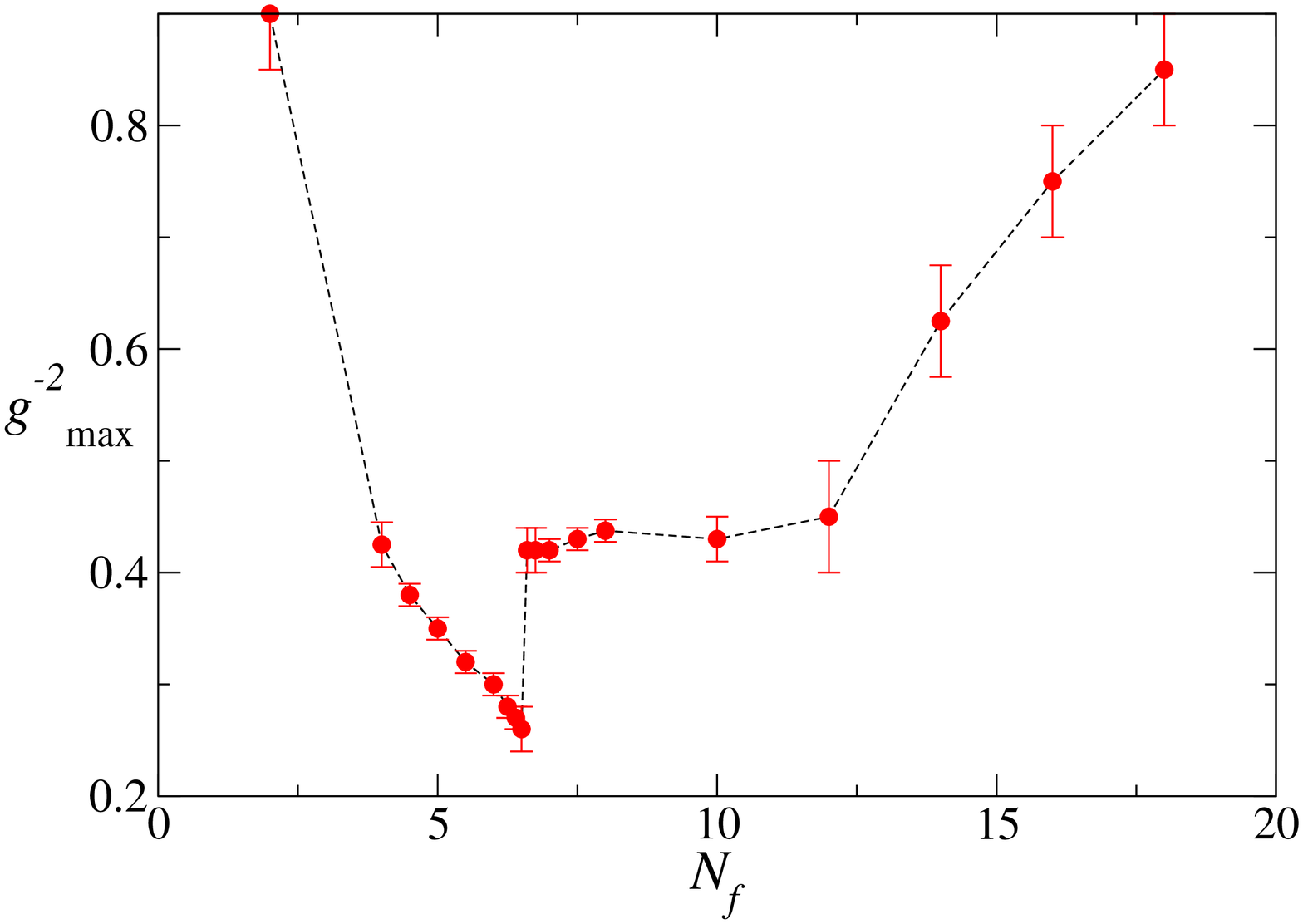}
\end{center}
\vspace{-.6cm}
\caption{
\label{fig:bmaxvsNf}
$g^2_{\rm max}$ vs. $N_f$.}
\vspace{-.25cm}
\end{figure}

Fig.~\ref{fig:psibpsivsb} also shows that for $N_f=8$ the peak moves to the 
right, and is considerably less pronounced. In Fig.~\ref{fig:bmaxvsNf}
we plot $g^{-2}_{\rm max}$ vs. $N_f$ for $N_f\in[2,18]$. For $N_f\lapprox6$,
$g^{-2}_{\rm max}$ decreases smoothly and monotonically, but in the interval
between 6.5 and 6.6 there is a sudden sharp increase; for $6.6\leq N_f
\leq12$, $g^{-2}\simeq0.43(2)$ is roughly constant, and then increases for
$N_f$ still larger. With the identification of $g^2_{\rm max}$ with the
strong coupling limit of the continuum theory, as argued above, it is natural
to ask whether the transition at $N_f\simeq6.5$ has any correlation with the 
chiral order parameter.
\begin{figure}[htb]
\begin{center}
\vspace*{-.8cm}
\epsfxsize=3.5in
\epsfbox{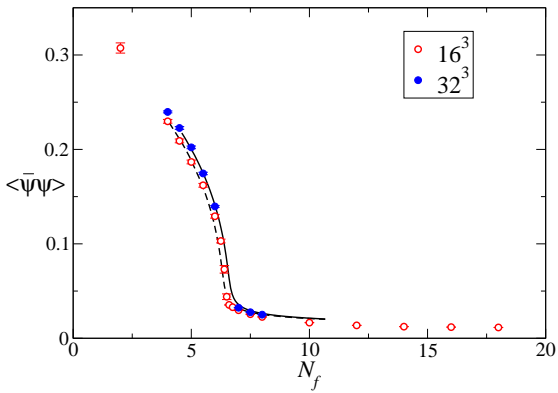}
\end{center}
\vspace{-.6cm}
\caption{
\label{fig:psibpsivsNf}
$\langle\bar\psi\psi(g^2_{\rm max})\rangle$ vs. $N_f$ for $m=0.01$. 
Lines denote
EoS fits discussed in the text.}
\vspace{-.25cm}
\end{figure}
Fig.~\ref{fig:psibpsivsNf} shows $\langle\bar\psi\psi(g^2_{\rm
max})\rangle$ vs. $N_f$. This plot necessarily puts an upper bound on the
condensate. It is difficult to draw any conclusion other than there being a
chiral symmetry restoring phase transition at 
\begin{equation}
N_{fc}=6.6(1).
\label{eq:Nfc}
\end{equation}
Eqn.~(\ref{eq:Nfc}) is the main result of this Letter.

In order to refine this picture,
further theoretical analysis is needed
to establish contact between the results shown in Fig.~\ref{fig:bmaxvsNf} and
the predictions of the $1/N_f$ expansion in the regime $N_f>N_{fc}$. In particular, 
there remains a disparity between $g^2_{\rm max}(N_f)$ 
and the large-$N_f$ prediction $g^2_{\rm
lim}={3\over2}$, which may be due to subleading corrections in $1/N_f$, or
to large finite-volume corrections described by 
the conformal field theory expected in the
limit $N_f\to N_{fc+}$. 

Fig.~\ref{fig:psibpsivsNf} is interesting because for the first time it
presents lattice data in a form suitable for direct comparison with SDE
predictions. One such study of chiral symmetry breaking in the 
$d$-dimensional Thirring model with $d\in(2,4)$ using this
approach was by Itoh
{\it et al\/} \cite{itoh}. They calculated the dressed fermion propagator
$S(p)=[A(p^2)p{\!\!\! /\,}+B(p^2)]^{-1}$, by first exploiting a hidden local 
symmetry to fix a gauge in which $A\equiv1$, and then
approximating the auxiliary propagator and fermion-auxiliary vertex by their
forms to leading order in $1/N_f$. This enabled a solution for the
self-energy function $\Sigma(x)=B(x)/\Lambda$, with $\Lambda$ a UV cutoff and
$x=(p/\Lambda)^{d-2}$, in the strong coupling limit $g^2\to\infty$, yielding  
a dynamically-generated fermion mass $M$:
\begin{equation}
\left({M\over\Lambda}\right)^{d-2}\propto
\exp\left({{-2\pi}\over\sqrt{{N_{fc}\over N_f}-1}}\right),
\label{eq:SDEsol}
\end{equation}
where $N_{fc}(d=3)=128/3\pi^2\simeq4.32$. The chiral order parameter 
follows via the relation
$\langle\bar\psi\psi\rangle\propto\Sigma^\prime(x=1)$, leading to the 
strong coupling prediction 
\begin{equation}
\langle\bar\psi\psi\rangle\propto\Lambda^{{d-2}\over2}M^{d\over2}.
\label{eq:SDEcond}
\end{equation}
Spontaneous chiral symmetry breaking occurs
for $N_f<N_{fc}$ in qualitative, but 
not quantitative agreement with Fig.~\ref{fig:psibpsivsNf}. 
Eqn.~(\ref{eq:SDEsol}) predicts an infinite-order phase transition. Its
nature  is further elucidated using
the anomalous scaling dimension
of the $\bar\psi\psi$ bilinear
$\gamma_{\bar\psi\psi}=d\ln\langle\bar\psi\psi\rangle/d\ln\Lambda$ and the 
relations for the critical exponents $\eta$ and $\delta$ \cite{annals}:
\begin{equation}
\eta=d-2\gamma_{\bar\psi\psi}\;;\;\;\;
\delta={{d+2-\eta}\over{d-2+\eta}}.
\end{equation}
From (\ref{eq:SDEcond}) we obtain $\gamma_{\bar\psi\psi}=(d-2)/2$ and hence
$\eta=2$, $\delta=1$. This scenario has been termed a ``conformal phase
transition'' \cite{conformal}, 
and has also been exposed by SDE approaches in QED$_3$ \cite{appelquist} and
quenched QED$_4$ \cite{miranskii}, 
the control parameter being respectively $N_f$ and the
fine structure constant $\alpha$.

The SDE analysis of \cite{itoh} was later extended to cover $g^2<\infty$
by Sugiura \cite{sugiura}. In this case the chiral transition takes place 
for $N_f<N_{fc}$, and the solution for the order parameter is of the form
\begin{equation}
\langle\bar\psi\psi\rangle\propto\Lambda^{d-2}M.
\end{equation}
The same chain of arguments leads to critical exponents
$\eta=4-d$, $\delta=d-1$, coincident with those of the $d$-dimensional
Gross-Neveu model in the large-$N_f$ limit \cite{annals}.
The nature of the transition predicted by SDEs
thus appears sensitive to the order of the 
limits $g^2\to\infty$, $N_f\to N_{fc}$.

We have been motivated by these considerations to attempt an EoS fit to the
data of Fig.~\ref{fig:psibpsivsNf} of the form 
\begin{equation}
m=A[(N_f-N_{fc})+CL^{-{1\over\nu}}]\langle\bar\psi\psi\rangle^p
+B\langle\bar\psi\psi\rangle^\delta,
\end{equation}
where $L$ is the linear extent of the system and the exponent $\nu$ is given by
the hyperscaling relation $\nu=(\delta+1)/d(\delta-p)$. The term proportional
to $C$ accounts for finite volume corrections. Our best fit to data with
$N_f\in[4,8]$ yields $N_{fc}=6.89(2)$, $\delta=6.90(3)$, $p=4.23(2)$, with 
$\chi^2$/dof=129/15. 

Besides the relatively poor quality, and significant
disagreement with
Eq.~(\ref{eq:Nfc}), the fit should not be
taken too seriously for two further reasons, namely the combination of ``exact''
HMC data with HMD containing a systematic error due to $\Delta\tau\not=0$, and
the as yet unquantified errors due to the identification of $g^2_{\rm lim}$ with
$g^2_{\rm max}$. Nonetheless the disparity of the value of the exponent $\delta$
with the SDE prediction is striking.
\begin{figure}[htb]
\begin{center}
\vspace*{-.8cm}
\epsfxsize=3.5in
\epsfbox{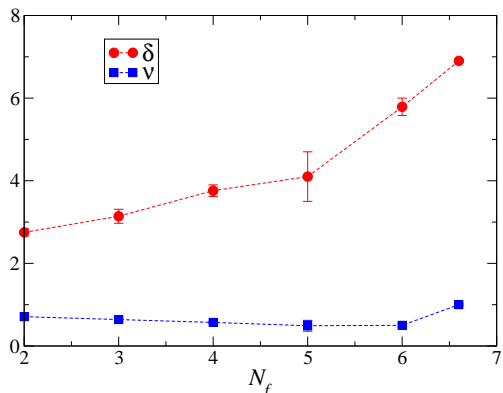}
\end{center}
\vspace{-1.0cm}
\caption{
\label{fig:exponents}
Critical exponents $\delta$ and $\nu$ vs. $N_f$.}
\vspace{-.25cm}
\end{figure}
Moreover, if we plot the fitted values for $\delta$ and $\nu$ together with 
those from EoS fits to data from $N_f<N_{fc}$ compiled from
Refs.~\cite{hands3,hands4,hands5}, as shown in Fig.~\ref{fig:exponents}, we see
that the estimate for $\delta(N_{fc})$ lies within a general trend of
$\delta$ increasing with $N_f$, with
no evidence for non-commutativity 
of the limits $g^2\to\infty$ and $N_f\to N_{fc}$ (there is some indication for 
a jump from $\nu\approx0.5$ to $\nu\approx1$ as $N_f$ increases from
6 to $N_{fc}$, which needs to be confirmed in a more refined study). 
In any case, the fitted values
of $\delta$ lie well above those predicted in the SDE approach.

In summary, for the first time we have been able using lattice Monte Carlo 
simulation to study a chiral phase
transition as the number of fermion flavors $N_f$ is varied continuously.
We find a continuous phase transition in the strong coupling limit at a critical
flavor number $N_{fc}=6.6(1)$, in qualitative, but {\em not} quantitative
agreement with the predictions of an analysis using Schwinger-Dyson equations.
In particular, the critical exponents are not those either of a conformal phase
transition or the 3$d$ Gross-Neveu model.
It is plausible that our result may inform estimates of the corresponding 
critical number of flavors for chiral symmetry breaking in QED$_3$, where
direct lattice simulations are hampered by a large separation of scales.
Note that a recent perturbative analysis of RG
flow equations in the large-$N_f$ limit of QED$_3$ predicts
$N_{fc}=6$~\cite{herbut2}.
In future it will be interesting to check whether the same universal features of
the strong coupling limit emerge using the alternative lattice formulations of the
Thirring model reviewed in \cite{hands3}.

\vspace{-12pt}

\acknowledgments

The simulations were performed on a cluster of 2.4GHz Opteron processors at 
the Frederick Institute of Technology, Cyprus.

\end{document}